\pdfoutput=1
\documentclass[twocolumn,aps]{revtex4}
\usepackage{graphicx}
\usepackage{amsmath}
\usepackage{amsfonts}
\begin{document}

%%%%%%%%%%%%%%%%%%%%%%%%%%%%%%%%%%%%%%%%%%%%%%%%%%%%%%%%
%% Kieron's DFT macros
%%%%%%%%%%%%%%%%%%%%%%%%%%%%%%%%%%%%%%%%%%%%%%%%%%%%%%%%

% Equation environments
\def\bea{\begin{eqnarray*}}
\def\eea{\end{eqnarray*}}
\def\ben{\begin{equation}}
\def\een{\end{equation}}
\def\benu{\begin{enumerate}}
\def\enu{\end{enumerate}}

% density
\def\n{n}

% Scriptstyle
\def\sss{\scriptscriptstyle\rm}

% gamma subscript for scaling
\def\g{_\gamma}

% lambda superscript for coupling constant
\def\l{^\lambda}
\def\lfc{^{\lambda=1}}
\def\lo{^{\lambda=0}}

% bits and pieces
\def\marnote#1{\marginpar{\tiny #1}}
\def\rsav{\langle r_s \rangle}
\def\invdif{\frac{1}{|\br_1 - \br_2|}}

%operators
\def\hatT{{\hat T}}
\def\hatV{{\hat V}}
\def\hatH{{\hat H}}
\def\1var{(\bx_1...\bx\N)}

% Fractions
\def\half{\frac{1}{2}}
\def\quart{\frac{1}{4}}

% Bold-face symbols
\def\bp{{\bf p}}
\def\br{{\bf r}}
\def\bR{{\bf R}}
\def\bu{{\bf u}}
\def\bx{{x}}
\def\by{{y}}
\def\ba{{\bf a}}
\def\bq{{\bf q}}
\def\bj{{\bf j}}
\def\bX{{\bf X}}
\def\bF{{\bf F}}
\def\bchi{{\bf \chi}}
\def\bof{{\bf f}}

% script symbols
\def\cA{{\cal A}}
\def\cB{{\cal B}}

% Standard subscripts
\def\x{_{\sss X}}
\def\c{_{\sss C}}
\def\s{_{\sss S}}
\def\xc{_{\sss XC}}
\def\dc{_{\sss DC}}
\def\Hxc{_{\sss HXC}}
\def\xj{_{{\sss X},j}}
\def\xcj{_{{\sss XC},j}}
\def\N{_{\sss N}}
\def\H{_{\sss H}}

% Word sub and superscripts
\def\ext{_{\rm ext}}
\def\pot{^{\rm pot}}
\def\hyb{^{\rm hyb}}
\def\hah{^{1/2\& 1/2}}
\def\LSD{^{\rm LSD}}
\def\LDA{^{\rm LDA}}
\def\GEA{^{\rm GEA}}
\def\GGA{^{\rm GGA}}
\def\SPL{^{\rm SPL}}
\def\sce{^{\rm SCE}}
\def\PBE{^{\rm PBE}}
\def\DFA{^{\rm DFA}}
\def\helm{^{\rm unamb}}
\def\una{^{\rm unamb}}
\def\ion{^{\rm ion}}
\def\I{^{\rm I}}
\def\pot{^{\rm pot}}
\def\sav{^{\rm sph. av.}}
\def\unif{^{\rm unif}}
\def\LSD{^{\rm LSD}}
\def\ee{_{\rm ee}}
\def\vir{^{\rm vir}}
\def\ALDA{^{\rm ALDA}}
\def\PGG{^{\rm PGG}}
\def\GK{^{\rm GK}}

% spin indices
\def\up{_\uparrow}
\def\dn{_\downarrow}
\def\up{_\alpha}
\def\dn{_\beta}

% Words
\def\td{time-dependent~}
\def\KS{Kohn-Sham~}
\def\DFT{density functional theory~}

%integrals
\def\fourint{ \int_{t_0}^{t_1} \! dt \int \! d^3r\ }
\def\fourintp{ \int_{t_0}^{t_1} \! dt' \int \! d^3r'\ }
\def\intx{\int\!d^4x}
\def\sph_int{ {\int d^3 r}}
\def\radint{ \int_0^\infty dr\ 4\pi r^2\ }

%journals
\def\PRA{Phys. Rev. A\ }
\def\PRB{Phys. Rev. B\ }
\def\PRL{Phys. Rev. Letts.\ }
\def\JCP{J. Chem. Phys.\ }
\def\JPCA{J. Phys. Chem. A\ }
\def\IJQC{Int. J. Quant. Chem.\ }

\title{Adiabatic Connection in the Low-Density Limit}
\author{Zhenfei Liu and Kieron Burke}
\affiliation{Department of Chemistry, University of California, Irvine, California, 92697-2025}
\date{\today}

\begin{abstract}
In density functional theory (DFT), the exchange-correlation functional can be exactly expressed by the adiabatic connection integral. It has been noticed that as $\lambda \to \infty$, the $\lambda^{-1}$ term in the expansion of $W(\lambda)$ vanishes. We provide a simple but rigorous derivation to this exact condition in this work. We propose a simple parametric form for the integrand, satisfying this condition, and show that it is highly accurate for weakly-correlated two-electron systems.
\end{abstract}

\maketitle

In density functional theory (DFT) \cite{FNM03}, the exchange-correlation functional $E\xc[n]$ is exactly expressed by the adiabatic connection \cite{LP75, GL76} formula:
\begin{equation}
E\xc[n]=\int_0^1 d\lambda \, W[n](\lambda),
\label{eq1}
\end{equation}
where $\lambda$ is a coupling constant that connects the Kohn-Sham system ($\lambda=0$) to the true system ($\lambda=1$), while keeping the density $n(\mathbf{r})$ fixed. The integrand, $W(\lambda)$, contains only potential contributions to $E\xc$. The shape of $W(\lambda)$ has been much studied in DFT \cite{PMTT08}. For example, the success of hybrid functionals that mix some fraction of exact exchange with a generalized gradient approximation (GGA) can be understood this way \cite{BEP97}. There is ongoing research to use the low density ($\lambda \to \infty$) limit as information in construction of accurate models of $W(\lambda)$ \cite{SPL99,SPK00,SPKb00}. Recently, the adiabatic connection formula has been used directly in functional construction \cite{MCY06}. 

The expansion of $W(\lambda)$ in the high-density (weak coupling) limit for finite systems is known to be \cite{SPK00}:
\begin{equation}
W(\lambda)=W_0+W_0' \lambda + \cdots \mbox{as $\lambda \to 0$},
\label{highd}
\end{equation}
where $W_0'=2E\c^{\rm{GL2}}$, with $E\c^{\rm{GL2}}$ the second-order coefficient in G\"{o}rling-Levy perturbation theory \cite{SPL99,GL93,GL95}. The expansion in the low-density (strongly correlated) limit is believed to be \cite{SPK00, GVS08}:
\begin{equation}
W(\lambda)=W_\infty+W_{\infty}' \lambda^{-1/2} + \cdots \mbox{as $\lambda \to \infty$},
\label{lowd}
\end{equation}
where $W_{\infty}'$ is defined as the coefficient of $\lambda^{-1/2}$ in the expansion above, and $W_{\infty}$ can be calculated from the strictly correlated electron (SCE) limit \cite{SGS07}. In addition to these expansions, by definition the exact $W[n](\lambda)$ is known to satisfy the following scaling property \cite{SPK00}:
\begin{equation}
W[n](\lambda)=\lambda W_1[n_{1/ \lambda}],
\label{scaling}
\end{equation}
where $n_{1/ \lambda}(\mathbf{r})$ is the scaled density, defined by $n_\gamma(\mathbf{r})=\gamma^3n(\gamma \mathbf{r})$, $0<\gamma<\infty$. In the equations above, one can show that $W_0=E\x$, the exchange energy, and that $W_\infty$ is finite \cite{SPL99}. The dependence on $\lambda^{-1/2}$ in the low-density limit is because correlation dominates here, and the Thomas-Fermi screening length is proportional to $\lambda_F^{-1/2}$.

In practical DFT calculations, $W(\lambda)$ must be approximated. However, any approximate $W(\lambda)$ should satisfy several exact conditions, such as Eqs. (\ref{highd}), (\ref{lowd}) and (\ref{scaling}).  In the erratum to Ref. \cite{SPK00}, Seidl et al. concluded that for the ISI model (see below), the spurious $\lambda^2 \ln \lambda$ term in $E\c[n_\lambda]$ is due to the $\lambda^{-1}$ term in the expansion of $W(\lambda)$ as $\lambda \to \infty$ [Eq. (\ref{lowd})]. In a recent work \cite{GVS08}, this was proved rigorously, but only by calculating zero-point oscillations about the strictly-correlated limit. In this paper, we provide a simple derivation and how this exact constraint affects approximate functionals. Throughout this paper, we use atomic units ($e^2=\hbar=\mu=1$) everywhere, i.e. all energies are in Hartrees and all distances in Bohr radii.

Any $\lambda$-dependence can always be expressed in terms of density scaling. Using the fundamental relation of Levy-Perdew equation \cite{LP85}, one finds:
\begin{equation}
W[n](\lambda)=E\x[n]-\gamma^2 \frac{d}{d\gamma} \left( \frac{E\c[n_\gamma]}{\gamma^2} \right),
\label{wlambda}
\end{equation}
and it is generally believed for nondegenerate Kohn-Sham systems \cite{UK02} that $E\c[n_\gamma]$ has the following expansion in the low density limit ($\gamma \to 0$):
\begin{equation}
E\c[n_\gamma]=\gamma \left( B_0[n]+\gamma^{1/2}B_1[n]+\gamma B_2[n]+\cdots \right),
\label{assume}
\end{equation}
where the $B_k[n]$'s ($k=0,1,2 \cdots$) are scale-invariant functionals. Substituting into Eq. (\ref{wlambda}), we find the expansion of $W(\lambda)$ for large $\lambda$:
\begin{equation}
W(\lambda)=E\x[n]+B_0[n]+\frac{1}{2}\lambda^{-1/2}B_1[n]-\frac{1}{2}\lambda^{-3/2}B_3[n]+\cdots
\label{wexpan}
\end{equation}
i.e. the $\lambda^{-1}$ term is missing, and $W(\lambda)$ is independent of $B_2[n]$. 

Now we survey approximations to $W(\lambda)$ and see whether they have the correct low-density expansion [Eq. (\ref{wexpan})]. There are several kinds of approximations, the most famous being the ISI (interaction-strength interpolation) model by Seidl et al \cite{SPL99,SPK00,SPKb00}:
\begin{equation}
W^{\rm {ISI}}[n](\lambda)=W_\infty[n]+\frac{X[n]}{\sqrt{1+Y[n]\lambda}+Z[n]},
\label{msisi}
\end{equation}
where $X=xy^2/z^2$, $Y=xX/z^2$, $Z=X/z-1,$ with $x=-2W_0'[n], y=W_\infty '[n],$ and $z=E\x[n]-W_\infty[n].$

The ISI model uses the values of $W[n]$ and its derivatives at both the high-density ($\lambda \to 0$) and the low-density ($\lambda \to \infty$) limits, to produce an accurate curve for $W(\lambda)$, $0 \le \lambda \le 1$, to insert in Eq. (\ref{eq1}) to get an approximation to $E\xc$. It gives very accurate results for the correlation energy \cite{SPK00} and meets several conditions. But if we expand $W^{\rm {ISI}}$ in the low density limit:
\begin{equation}
W^{\rm {ISI}}(\lambda)=W_\infty+\frac{X}{\sqrt{Y}} \lambda^{-1/2}+\frac{XZ}{Y} \lambda^{-1} + \cdots,
\end{equation}
we can see that its $\lambda^{-1}$ term does not generally vanish, although it works very well numerically for $E\c$ \cite{PKS01}. This wrong coefficient was already shown to produce a spurious term $(\lambda^2 \ln \lambda)$ in the expansion of $E\c[n_\lambda]$ as $\lambda \to \infty$ \cite{SPK00}.

There were several attempts to overcome this problem [correctly omitting the $\lambda^{-1}$ term but including all the other (integer and half-integer powers) terms] in the literature \cite{S07,GVS08} by modifying the ISI model, but they are less simple: one requires $W_0''$ [the next order in Eq. (\ref{highd})] \cite{S07} and the other is not a direct model to $W_\lambda$ \cite{GVS08}. Consider instead the following 4-parameter interpolation model:
\begin{equation}
W^{\rm acc}(\lambda)=a+by+dy^4, \hspace{0.3in} y=\frac{1}{\sqrt{1+c\lambda}},
\label{mod4}
\end{equation}
where $a,b,c,$ and $d$ are scale-invariant functionals. We use the same inputs as those for the ISI model, i.e. $W_0, W_0', W_\infty$, and $W_\infty'$, to fit the parameters. Generally there are no analytical expressions in compact form for the parameters, and one has to solve for them numerically. The 4th power in $y$ is the lowest that can be added while satisfying the exact conditions, but producing an expansion with non-zero $\lambda^{-n}$ terms ($n \in \mathbb{Z}, n > 1$). We recommend use of this $W^{\rm acc}$ to replace the ISI model because it is numerically accurate and avoids the $\lambda^{-1}$ term in the low-density limit. One can show that $W^{\rm acc}$ obeys the scaling property [Eq. (\ref{scaling})],  provided that $W_0[n_\gamma]=\gamma W_0[n]$, $W_0'[n_\gamma]=W_0'[n]$, $W_\infty[n_\gamma]=\gamma W_\infty[n]$, and $W_\infty'[n_\gamma]=\gamma^{3/2} W_\infty'[n]$, as they should. If we integrate $W^{\rm acc}(\lambda)$ over $\lambda$ from 0 to 1, we find a simple expression for the exchange-correlation energy:
\begin{equation}
E\xc^{\rm acc}=a+\frac{d}{1+c}+2b(-1+\sqrt{1+c})/c.
\label{ecmod4}
\end{equation}

We compare the performance of the new model and ISI on Hooke's atom, two electrons in a spherical harmonic well, with force constant $k=1/4$. We show below that for this system, our $W^{\rm acc}$ works as a highly-accurate interpolation, even more accurate than the ISI model.

\begin{figure}[h]
\begin{center}
\includegraphics[width=3.5in]{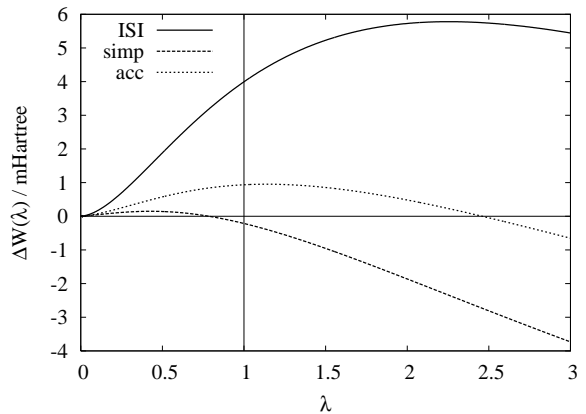}
\caption{Comparison of three different approximations to $W(\lambda)$ for Hooke's atom ($k=1/4$), plotted as $\Delta W=W^{\rm model}-W^{\rm exact}$. The exact curve (up to $\lambda=3$) is taken from Ref. \cite{MTB03}.}
\label{fghooke}
\end{center}
\end{figure}

Magyar et al. \cite{MTB03} calculated the $W(\lambda)$ curve for $0 \leq \lambda \leq 4$ for Hooke's atom ($k=1/4$) using $W_0=E\x=-0.515$ and $W_0'=-0.101$ as inputs. They confirmed that $W_\infty=-0.743$, consistent with the SCE ansatz \cite{SPL99}. They also found $W_\infty'=0.235$, but this was based on a fit that violated our condition, so we discount this result. Gori-Giorgi \cite{P09} calculated $W_\infty'=0.208$ based on the SCE model \cite{SPL99, GVS08}, which we consider exact. We apply these inputs ($W_0, W_0', W_\infty$, and $W_\infty'$) to our $W^{\rm acc}$ and the ISI model ($W^{\rm acc}$ generates two sets of solutions for $a,b,c,$ and $d$, but we select the one with $d$ closest to $b$, for it can be reduced to $W^{\rm simp}$ as below). We plot the differences between these models and the exact curve (taken from Ref. \cite{MTB03}) in Fig. \ref{fghooke}. One can see that our $W^{\rm acc}$ works very well between $\lambda=0$ and $1$, which is the range of interest. Its predictions for $W_1', E\c$, and $E\c+T\c$ are excellent, with $T\c$ being the correlation energy from the kinetic part, as listed in Table \ref{tbhooke}. With these exact inputs, we found that, as $\lambda \to \infty$, $W^{\rm ISI} \to -0.743+0.208\lambda^{-1/2}+0.068\lambda^{-1}+\cdots$, which shows that although the coefficient of $\lambda^{-1}$ is small, it does not vanish.

\begin{table}[h]
\caption{Comparison of several quantities for three different approximations to $W(\lambda)$ for Hooke's atom ($k=1/4$). The exact values are taken from Ref. \cite{MTB03} except for $W_\infty'$ \cite{P09}. All energies are in mHartrees.}
\begin{center}
\begin{tabular}{c|c c c c}
\hline\hline
 & exact & ISI & simp & acc\\
\hline
$W_1$ & -583 & -579 & -583 & -582\\
$W_1'$ & -44 & -41 & -45 & -44\\
$E\c$ & -39 & -37 & -38 & -38\\
$E\c+T\c$ & -10 & -10 & -9 & -9\\
\hline\hline
\end{tabular}
\end{center}
\label{tbhooke}
\end{table}

\begin{figure}[h]
\begin{center}
\includegraphics[width=3.5in]{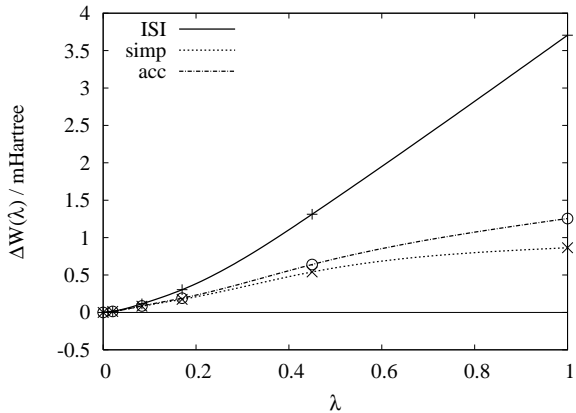}
\caption{Comparison of three different approximations to $W(\lambda)$ for helium atom, plotted as $\Delta W=W^{\rm model}-W^{\rm exact}$. The discrete values are shown, as well as fitting curves to aid the eyes. $W^{\rm exact}$ values (up to $\lambda=1$) are taken from Ref. \cite{FTB00}.}
\label{fghe}
\end{center}
\end{figure}

We can also apply our $W^{\rm acc}$ to the helium atom. Here $W_0=E\x=-1.025$, $W_0'=-0.095$ \cite{CS99}, and $W_\infty=-1.500$ \cite{SPL99}, $W_\infty'=0.621$ \cite{GVS08} from the SCE model \cite{SPL99, GVS08}. We plot the differences between these models and the exact curve (taken from Ref. \cite{FTB00}) in Fig. \ref{fghe} and compare several key quantities in Table \ref{tbhe}.

One can see that our model here works fairly well, and $W^{\rm simp}$ (see below) is even a little better than $W^{\rm acc}$. ISI does not satisfy the exact condition we derived in this work [Eq. (\ref{wexpan})]: as $\lambda \to \infty$, $W^{\rm ISI} \to -1.500+0.621 \lambda^{-1/2}+0.376 \lambda^{-1}+\cdots$, so the $\lambda^{-1}$ coefficient is not even small.

\begin{table}[h]
\caption{Comparison of several quantities for three different approximations to $W(\lambda)$ for helium atom. The exact values are taken from Ref. \cite{FTB00}. All energies are in mHartrees.}
\begin{center}
\begin{tabular}{c|c c c c}
\hline\hline
 & exact & ISI & simp & acc\\
\hline
$W_1$ & -1104 & -1100 & -1103 & -1103\\
$W_1'$ & -64 & -60 & -64 & -63\\
$E\c$ & -42 & -40 & -42 & -41\\
$E\c+T\c$ & -6 & -6 & -5 & -5\\
\hline\hline
\end{tabular}
\end{center}
\label{tbhe}
\end{table}

Now, we propose a simpler version of $W^{\rm acc}$, which cannot be used in typical cases, as the exact value of $W_\infty'$ is not known in general. A simpler model is constructed by setting $d=b$, to yield:
\begin{equation}
W^{\rm simp}(\lambda)=a+b(y+y^4), \hspace{0.3in} y=\frac{1}{\sqrt{1+c\lambda}},
\label{easy2}
\end{equation}
with $a, b$ and $c$ being scale-invariant functionals. We have found (see results for Hooke's atom and helium atom) that although there is one parameter less, the above form produces usefully accurate results, especially between $\lambda=0$ and $1$. In a word, $W^{\rm acc}$ acts as an accurate interpolation to the whole adiabatic connection curve, while $W^{\rm simp}$ is more convenient and practical to use, without losing accuracy. It yields $W_\infty'=0.191$ for Hooke's atom and $0.594$ for helium.

We use $W_0$, $W_\infty$ and $W_0'$ to construct the explicit form of $W^{\rm simp}(\lambda)$, and find:
\begin{equation}
a=W_\infty, \hspace{0.1in} b=\frac{W_0-W_\infty}{2}, \hspace{0.1in} c=\frac{4W_0'}{5(W_\infty-W_0)}.
\label{para}
\end{equation}
Thus $a$ and $b$ set the endpoints, while $c$ is a measure of the curvature. Substituting Eq. (\ref{para}) into Eq. (\ref{easy2}), we get the explicit form of $W(\lambda)$ in terms of $W_0$, $W_\infty$ and $W_0'$. One can show that it has the correct expansion in both limits, and it obeys the scaling property [Eq. (\ref{scaling})]. Setting $d=b$ in Eq. (\ref{ecmod4}) and subtracting exchange, it yields:
\begin{equation}
E\c^{\rm simp}=2b[f(c)-1], \hspace{0.1in} f(c)=\lbrack \sqrt{1+c}-\frac{1+c/2}{1+c} \rbrack /c,
\label{ecmod3}
\end{equation}
with $b$ and $c$ defined in Eq. (\ref{para}). $E\c^{\rm simp}$ correctly recovers GL2 in the weakly-correlated limit ($W_\infty \to -\infty$, keeping $W_0$ and $W_0'$ fixed, such as in the $Z \to \infty$ limit of two-elecron ions) and $E\xc^{\rm simp}$ correctly reduces to $W_\infty$ for strong static correlation ($W_0' \to -\infty$, keeping $W_0$ and $W_\infty$ fixed, such as for stretched H$_2$). We can calculate the kinetic correlation energy $T\c$:
\begin{equation}
T\c=b[2f(c)-z-z^4],
\end{equation}
with $f(c)$ defined in Eq. (\ref{ecmod3}) and $z=1/\sqrt{1+c}$, showing that the curvature $\beta=T\c/|E\c-T\c|$ \cite{BPEb97} is a function of $c$ alone. We strongly urge $E\xc^{\rm simp}$ be applied whenever its inputs are accurately known.

We can further test our $W^{\rm simp}$ in systems with more than two electrons, but only those for which all inputs are known, with results listed in Table \ref{tbmore2}. One can see that $W^{\rm simp}$ predicts $E\c$ fairly accurately, but is less accurate than $W^{\rm ISI}$. This is perhaps due to lack of $W_\infty'$ in $W^{\rm simp}$.

\begin{table}[h]
\caption{Comparison of $W^{\rm simp}$ and $W^{\rm ISI}$ on systems with more than two electrons. $E\x$, $W_0'$ and $W_\infty$ are taken from Ref. \cite{SGS07}, and $W_\infty'$ is taken from Ref. \cite{GVS08}. All energies are in Hartrees.}
\begin{center}
\begin{tabular}{cccccccc}
\hline\hline
 & $E\x$ & $W_0'$ & $W_\infty^{\rm SCE}$ & $W_\infty'^{\rm SCE}$ & $E\c^{\rm ISI}$ & $E\c^{\rm simp}$ & $E\c^{\rm exact}$ \\
\hline
Be & -2.67 & -0.250 & -4.02 & 2.59 & -0.104 & -0.110 & -0.096 \\
Ne & -12.1 & -0.938 & -20.0 & 22.0 & -0.410 & -0.432 & -0.394 \\
\hline\hline
\end{tabular}
\end{center}
\label{tbmore2}
\end{table}

In fact, in their first paper on the ISI model, Seidl et al. proposed a similar model \cite{SPL99}, which yields results numerically very close to those of ISI, but without the $y^4$ term. But their model contains no $\lambda^{-n} (n>1)$ contributions. Note that none of these models work for the uniform electron gas, because $W_0'=-\infty$ \cite{PKS01}, so both the model developed by Seidl et al. \cite{SPL99} and $W^{\rm simp}$ reduce to $W(\lambda)=W_\infty$.

After the bulk of this work was completed, we received a preprint of Ref. \cite{GVS08}, containing a detailed theory of the leading corrections to $W(\lambda)$ as $\lambda \to \infty$, consistent with the much simpler arguments given here. Also, we use their $W_\infty'$ value for helium (see text) to replace the old one predicted by point-charge-plus-continuum (PC) model \cite{SPK00}.

We thank professor John Perdew for kind discussions. This work is supported by National Science Foundation under grant number CHE-0809859.

\end{document}